
\overfullrule=0pt
\def\singlespace{\normalbaselines}
\def\oneandahalfspace{\baselineskip=16pt plus 1pt
\lineskip=2pt\lineskiplimit=1pt}

\def\np{\vfill\eject}
\def\nl{\hfil\break}

\def\nofirstpagenoten{\nopagenumbers\footline={\ifnum\pageno>1\tenrm
\hss\folio\hss\fi}}
\def\nofirstpagenotwelve{\nopagenumbers\footline={\ifnum\pageno>1\twelverm
\hss\folio\hss\fi}}
\def\leaderfill{\leaders\hbox to 1em{\hss.\hss}\hfill}


\parindent=20pt
\def\narrow{\advance\leftskip by 40pt \advance\rightskip by 40pt}
\baselineskip=20pt
\def\AB{\bigskip
        \centerline{\bf ABSTRACT}\medskip\narrow}
\def\nonarrower{\advance\leftskip by -40pt\advance\rightskip by -40pt}
\def\AE{\bigskip\nonarrower}

\def\boxit#1{\vbox{\hrule\hbox{\vrule\kern3pt
        \vbox{\kern3pt#1\kern3pt}\kern3pt\vrule}\hrule}}

\def\gtorder{\mathrel{\raise.3ex\hbox{$>$}\mkern-14mu
             \lower0.6ex\hbox{$\sim$}}}
\def\ltorder{\mathrel{\raise.3ex\hbox{$<$}|mkern-14mu
             \lower0.6ex\hbox{\sim$}}}
\def\dalemb#1#2{{\vbox{\hrule height .#2pt
        \hbox{\vrule width.#2pt height#1pt \kern#1pt
                \vrule width.#2pt}
        \hrule height.#2pt}}}

\font\fourteentt=cmtt12 scaled \magstep1\font\fourteenbf=cmbx12 scaled
\magstep1
\font\fourteenrm=cmr12 scaled \magstep1\font\fourteeni=cmmi12 scaled \magstep1
\font\fourteenss=cmss12 scaled \magstep1
\font\fourteensy=cmsy10 scaled \magstep2 \font\fourteensl=cmsl12 scaled
\magstep
1
\font\fourteenex=cmex10 scaled \magstep2 \font\fourteenit=cmti12 scaled
\magstep
1
\font\twelvett=cmtt12 \font\twelvebf=cmbx12
\font\twelverm=cmr12 \font\twelvei=cmmi12 \font\twelvess=cmss12
\font\twelvesy=cmsy10 scaled \magstep1 \font\twelvesl=cmsl12
\font\twelveex=cmex10 scaled \magstep1 \font\twelveit=cmti12
\font\tenss=cmss10
 
 \font\ninebf=cmbx9
\font\ninerm=cmr9 \font\ninei=cmmi9
\font\ninesy=cmsy9 
\font\eightrm=cmr8
\catcode`@=11
\newskip\ttglue
\newfam\ssfam

\def\fourteenpoint{\def\rm{\fam0\fourteenrm}
\textfont0=\fourteenrm \scriptfont0=\tenrm \scriptscriptfont0=\sevenrm
\textfont1=\fourteeni \scriptfont1=\teni \scriptscriptfont1=\seveni
\textfont2=\fourteensy \scriptfont2=\tensy \scriptscriptfont2=\sevensy
\textfont3=\fourteenex \scriptfont3=\fourteenex \scriptscriptfont3=\fourteenex
\def\it{\fam\itfam\fourteenit} \textfont\itfam=\fourteenit
\def\sl{\fam\slfam\fourteensl} \textfont\slfam=\fourteensl
\def\bf{\fam\bffam\fourteenbf} \textfont\bffam=\fourteenbf
\scriptfont\bffam=\tenbf \scriptscriptfont\bffam=\sevenbf
\def\tt{\fam\ttfam\fourteentt} \textfont\ttfam=\fourteentt
\def\ss{\fam\ssfam\fourteenss} \textfont\ssfam=\fourteenss
\tt \ttglue=.5em plus .25em minus .15em
\normalbaselineskip=16pt
\abovedisplayskip=16pt plus 3pt minus 10pt
\belowdisplayskip=16pt plus 3pt minus 10pt
\abovedisplayshortskip=0pt plus 3pt
\belowdisplayshortskip=8pt plus 3pt minus 5pt
\parskip=3pt plus 1.5pt
\setbox\strutbox=\hbox{\vrule height10pt depth4pt width0pt}
\let\sc=\tenrm
\let\big=\fourteenbig \normalbaselines\rm}
\def\fourteenbig#1{{\hbox{$\left#1\vbox to10pt{}\right.\n@space$}}}

\def\twelvepoint{\def\rm{\fam0\twelverm}
\textfont0=\twelverm \scriptfont0=\ninerm \scriptscriptfont0=\sevenrm
\textfont1=\twelvei \scriptfont1=\ninei \scriptscriptfont1=\seveni
\textfont2=\twelvesy \scriptfont2=\ninesy \scriptscriptfont2=\sevensy
\textfont3=\twelveex \scriptfont3=\twelveex \scriptscriptfont3=\twelveex
\def\it{\fam\itfam\twelveit} \textfont\itfam=\twelveit
\def\sl{\fam\slfam\twelvesl} \textfont\slfam=\twelvesl
\def\bf{\fam\bffam\twelvebf} \textfont\bffam=\twelvebf
\scriptfont\bffam=\ninebf \scriptscriptfont\bffam=\sevenbf
\def\tt{\fam\ttfam\twelvett} \textfont\ttfam=\twelvett
\def\ss{\fam\ssfam\twelvess} \textfont\ssfam=\twelvess
\tt \ttglue=.5em plus .25em minus .15em
\normalbaselineskip=14pt
\abovedisplayskip=14pt plus 3pt minus 10pt
\belowdisplayskip=14pt plus 3pt minus 10pt
\abovedisplayshortskip=0pt plus 3pt
\belowdisplayshortskip=8pt plus 3pt minus 5pt
\parskip=3pt plus 1.5pt
\setbox\strutbox=\hbox{\vrule height10pt depth4pt width0pt}
\let\sc=\ninerm
\let\big=\twelvebig \normalbaselines\rm}
\def\twelvebig#1{{\hbox{$\left#1\vbox to10pt{}\right.\n@space$}}}

\def\tenpoint{\def\rm{\fam0\tenrm}
\textfont0=\tenrm \scriptfont0=\sevenrm \scriptscriptfont0=\fiverm
\textfont1=\teni \scriptfont1=\seveni \scriptscriptfont1=\fivei
\textfont2=\tensy \scriptfont2=\sevensy \scriptscriptfont2=\fivesy
\textfont3=\tenex \scriptfont3=\tenex \scriptscriptfont3=\tenex
\def\it{\fam\itfam\tenit} \textfont\itfam=\tenit
\def\sl{\fam\slfam\tensl} \textfont\slfam=\tensl
\def\bf{\fam\bffam\tenbf} \textfont\bffam=\tenbf
\scriptfont\bffam=\sevenbf \scriptscriptfont\bffam=\fivebf
\def\tt{\fam\ttfam\tentt} \textfont\ttfam=\tentt
\def\ss{\fam\ssfam\tenss} \textfont\ssfam=\tenss
\tt \ttglue=.5em plus .25em minus .15em
\normalbaselineskip=12pt
\abovedisplayskip=12pt plus 3pt minus 9pt
\belowdisplayskip=12pt plus 3pt minus 9pt
\abovedisplayshortskip=0pt plus 3pt
\belowdisplayshortskip=7pt plus 3pt minus 4pt
\parskip=0.0pt plus 1.0pt
\setbox\strutbox=\hbox{\vrule height8.5pt depth3.5pt width0pt}
\let\sc=\eightrm
\let\big=\tenbig \normalbaselines\rm}
\def\tenbig#1{{\hbox{$\left#1\vbox to8.5pt{}\right.\n@space$}}}
\let\rawfootnote=\footnote \def\footnote#1#2{{\rm\parskip=0pt\rawfootnote{#1}
{#2\hfill\vrule height 0pt depth 6pt width 0pt}}}
\def\tenfoot{\tenpoint\hskip-\parindent\hskip-.1cm}

\twelvepoint
\def\sbullet{\raise.2em\hbox{$\scriptscriptstyle\bullet$}}
\nofirstpagenotwelve
\baselineskip 15pt

\baselineskip=12pt
\rightline{UR--1232}
\rightline{ER--13065--685}
\rightline{CTP TAMU--54/91}
\rightline{IC/91/383 }
\rightline{UFIFT--HEP--91--26}
\rightline{November 1991}

\vskip 1truecm
\centerline{\bf The Super $W_\infty$ Symmetry of the Manin-Radul Super KP
Hierarchy}
\vskip
1.5truecm \centerline{A. Das,$^1$\footnote{$^{\star}$}{Supported in
 part by the
U.S. Department of Energy, under  grant DE-AC02-76ER-13065.}
E. Sezgin,$^2$\footnote{$^{\dag}$}{Supported in part by the U.S.
National Science Foundation, under grant
PHY-9106593.} and S.J. Sin $^3$\footnote{$^{\dag\dag}$}
{Supported in part by the U.S. Department of Energy, under  grant
DE-FG05-86ER-40272.}}
\vskip
1.5truecm  \item{$^1$}{\it Department of Physics and Astronomy,
University of Rochester, Rochester, NY 14627, USA}
\item{$^2$}{\it Center for
Theoretical Physics, Texas A\&M University, College Station,\nl TX 77843--4242,
USA\/}
\item{$^3$}{\it Department of Physics, University of Florida,
Gainesville, FL 32611, USA} \vskip 1.5truecm
\AB\singlespace
      We show that the Manin-Radul super KP hierarchy is invariant under
super $W_\infty$
transformations. These transformations are characterized by time dependent
flows  which commute with
the usual flows generated by the conserved quantities of the super KP
hierarchy.
 \AE\oneandahalfspace  \np

\noindent
{\bf 1. Introduction}
\bigskip

There has been a great deal of interest in the study of conformal symmetries in
the past several years. In the context of high energy physics, the motivation
for such studies came from string theories and has already led to some
remarkable results. The conformal symmetries, in turn, have led in a natural
way to the study of $W_N$ algebras [1]. While $W_N$ algebras do not define a
Lie
algebra, $W_\infty$ or $W_{1+\infty}$ algebras [2] are proper Lie algebras
and are
expected to play a role in the understanding of string theories [3][4].

In a completely parallel development, the integrable models have also attracted
a lot of attention in recent years. For example, it is known now that one of
the Hamiltonian structures (Poisson brackets) of the KdV
 equation is nothing other than
the Virasoro algebra [5]. Similarly, it has become clear that the $W_N$
algebras
arise as Hamiltonian structures of various integrable models in 1+1
dimensions [6].  The integrable models also play an important role in the
study of various 2D gravity theories. For example, the gravitational Ward
identity for pure 2D gravity in the light-cone gauge turns out to be none
other than the KdV hierarchy equation [7] and similarly for other gravity
theories.

The KP (Kadomtsev-Petviashvili) equation [8] is a 2+1 dimensional integrable
system which leads to a large class of 1+1 dimensional integrable models upon
appropriate reduction. In terms of the dynamical variable $u(x,y,t)$, the KP
equation reads
$$
  {\partial\over \partial x} \bigg({\partial u\over \partial t} -{1\over 4}
{\partial^3 u\over \partial x^3}-3u{\partial u\over \partial x}\bigg) = {3\over
4}
{\partial^2 u \over \partial y^2}  \eqno(1.1)
$$
As is obvious, in the absence of y-dependence, Eq. (1.1) reduces to the KdV
equation which, we have argued, plays an important role in the study of
strings. It is, therefore, natural to expect that the KP equation may
provide further understanding of the various string theories. In fact, it
has already been pointed out that one of the Hamiltonian structures of the
KP equation is isomorphic to the $W_{1+\infty}$ algebra [9] which
is expected to play a significant role in the understanding of strings. It
has also been argued that the KP hierarchy admits symmetries which satisfy
the $W_{1+\infty}$ algebra [10][3].

One can, of course, supersymmetrize various integrable systems. In fact, the
KP equation allows for more than one supersymmetric generalization [11][12].
And it
is, of course, ultimately the superstring theory which is of physical
interest. Therefore, it is the structure of the supersymmetries and the
algebra of the symmetries for the supersymmetric systems that will be of
direct physical significance. It is with this goal that we have undertaken,
in this paper, the study of symmetries for the simplest of the super KP
hierarchies, namely, the Manin-Radul hierarchy [11]. In Sec. II, we describe
our
notation and recapitulate, very briefly, the essential features of the
Manin-Radul hierarchy. The symmetry conditions are discussed in detail in Sec.
III and explicit symmetry generators as well as the symmetry
transformations for the Manin-Radul hierarchy are given in Sec. IV. We
would like to emphasize here that the usual conserved quantities associated
with an integrable system, of course, generate symmetries of the system.
But what we are interested in are additional symmetries--in general time
dependent. We find an infinite set of bosonic and fermionic generators of
symmetry for the super KP hierarchy. In Sec. V, we show that the algebra
generated by these generators is precisely the super $W_\infty$. Our
conclusions are presented in Sec. VI.
\bigskip
\noindent
{\bf 2. The Manin-Radul Super KP Hierarchy}
\bigskip
The supersymmetric extension of the KP hierarchy
 introduced by Manin and Radul [11] is a system of
nonlinear equations for an infinite set of even and odd functions, depending on
a pair of odd and even space variables $(\xi,x)$ and  the odd-even
 times $(\tau_1,t_2, \tau_3, t_4,...)$ \footnote{$^\dagger$}{\tenfoot $t^2$
and $t^4$ are to be
identified with y and t, respectively. In the bosonic case, the KP equation
(1.1) then arises as
the lowest member of a hierarchy of equations.}.
 The manifold on which the solutions are defined, in this case, is a graded
manifold.  On this
manifold, we can, of course, define the usual supercovariant derivative
$$
\theta = {\partial \over
\partial \xi} + \xi \>
{\partial \over \partial x} \eqno(2.1)
$$
which satisfies
$$
[\theta, \theta] = 2 \> {\partial \over \partial x} \eqno(2.2)
$$
where the graded commutator is defined by $[A,B] = A B - (-1)^{ab} BA$ with
$a,b$ denoting the gradings of $A$ and $B$ and taking values 0 and 1
depending on whether the variable is bosonic or fermionic.  A formal
inverse of $\theta$ is defined to be
$$
\theta^{-1} = \xi + {\partial \over \partial \xi} \left({\partial
 \over \partial x} \right)^{-1} \eqno(2.3)
$$

On this manifold, we can also define the even and odd time derivatives as

$$
\eqalign{\theta_{2i} &= {\partial \over \partial t_{2i}}\cr
&\qquad\qquad\qquad\qquad\qquad \qquad\qquad i,j = 1,2, \dots\cr
\noalign{\vskip 4pt}%
\theta_{2i-1} &= {\partial \over \partial \tau_{2i-1}} +
\sum^\infty_{j=1} \tau_{2j-1} \>
{\partial \over \partial t_{2i+2j-2}}\cr}\eqno(2.4)
$$
These time derivatives satisfy the algebra
$$
\eqalign{\big[ \theta_{2i}, \theta_{2j} \big] &= 0\cr
\big[ \theta_{2i}, \theta_{2j-1} \big] &= 0\cr
\big[ \theta_{2i-1}, \theta_{2j-1} \big] &=
 2 \theta_{2i+2j-2}\cr}\eqno(2.5)
$$
Furthermore, it is easy to see that both the even and the odd time
derivatives have vanishing commutator with $\theta$, namely,
$$
\left[ \theta , \theta_{2i} \right] = 0 =
\left[ \theta, \theta_{2i-1} \right] \eqno(2.6)
$$
In other words, these derivatives are covariant with respect to
supersymmetry transformations.

With these preliminaries, we can define the Lax operator for the
Manin-Radul super KP hierarchy as a pseudo-differential operator on this
graded manifold with the form
$$
L = \theta + \sum^\infty_{i=1} U_i \theta^{-i} \eqno(2.7)
$$
where the $U_i$'s are functions of all the even and odd variables with
the grading $(i +1)$.  This is completely parallel to the bosonic KP
hierarchy where the Lax operator
is defined as a pseudo-differential operator involving
${\partial \over \partial x}$.  The generalized Leibnitz rule for the
supercovariant
derivatives is given by
$$
\theta^i U = \sum^\infty_{j=0} (-1)^{u(i-j)} \pmatrix{i\cr
j\cr} \left( \theta^j U \right) \theta^{i-j} \eqno(2.8)
$$
where the super-binomial coefficients are defined for $i \geq 0$ by
$$
\pmatrix{i\cr
j\cr} = \cases{0 &for $j < 0 \> {\rm or}\>
j> i \> {\rm or} \> (i,j) = (0,1) \> {\rm mod}\> 2$\cr
\noalign{\vskip 8pt}%
\pmatrix{\big[ {i \over 2} \big]\cr
\noalign{\vskip 4pt}%
\big[ {j \over 2} \big]\cr} &for $0 \leq j \leq i \> {\rm and}\>
(i,j) \not= (0,1)\> {\rm mod}\> 2$\cr}\eqno(2.9)
$$
For $i < 0$, we define
$$
\pmatrix{i\cr j\cr} = (-1)^{[{j \over 2}]} \pmatrix{i + j-1\cr j\cr}\eqno(2.10)
$$

The Manin-Radul super KP hierarchy can, then, be described in terms of the Lax
equations [11]
$$
\theta_i L = \left[ \left(L^i \right)_-,L \right] \qquad\qquad i = 1,2,\dots
\eqno(2.11)
$$
where by $A_+, A_-$, we will understand the parts of the pseudo differential
operator $A$
containing nonnegative and only negative powers of $\theta$ respectively.
The structure of Eq. (2.11) is quite analogous to the Lax equation for the
bosonic KP hierarchy.  Let us note here that the equations in Eq. (2.11)
can also be written in an equivalent alternate form as
$$
\eqalign{\theta_{2i} &= - \big[ \big( L^{2i} \big)_+, L \big]\cr
\theta_{2i-1} &= - \big[ \big( L^{2i-1} \big)_+ , L \big]
 + 2 L^{2i} \cr}\eqno(2.12)
$$
Furthermore, we can introduce a dressing operator, as in the bosonic case,
by
$$
L = K \theta K^{-1} \eqno(2.13)
$$
with
$$
K = 1+ \sum^\infty_{i=1} K_i \theta^{-i}. \eqno(2.14)
$$
Consistency would then require that $K_1+{1\over 2}K_0=0.$ The Lax equation
(2.11) can now be
written as
$$
\theta_i K = \left( L^i \right)_- K \eqno(2.15)
$$

With a little bit of algebra, one can easily show that the Lax equation
(2.11) or equivalent Eq.(2.15) are consistent with the algebraic structure in
Eq.(2.5).  For the even time derivatives, then, we obtain the zero curvature
condition
$$
\theta_{2i} \left( L^{2j} \right)_- -
\theta_{2j} \left( L^{2i} \right)_- -
 \left[ \left( L^{2i} \right)_-, \left( L^{2j} \right)_- \right]= 0.
\eqno(2.16)
$$
\bigskip
\noindent{\bf 3. The Symmetries of the Manin-Radul Super KP Hierarchy}
\bigskip
The symmetries of the super KP hierarchy can be discussed equivalently at
various levels.  For example, one can study the symmetries of the linear
equation or the symmetries of the evolution of the dressing operator or the
symmetries of the Lax equation.  From our point of view, it is most useful
to study the evolution of the dressing operator.  We know that
(see Eq. (2.15))
$$
\theta_i K = \left( L^i \right)_- K \eqno(3.1)
$$
Consider an infinitesimal deformation of the dressing operator $K$, such that
$$
\delta K = \epsilon Q_- K \eqno(3.2)
$$
where $\epsilon$ is an infinitesimal constant parameter of deformation.
Note that we have restricted the operator Q to its negative part. As we shall
see, this
restriction turns out to simplify considerably the symmetry condition on Q
which we derive below.
One can think of the deformation as being generated through a flow equation
$$
\phi K = Q_- K \eqno(3.3)
$$
This deformation will be a symmetry of the super KP equation (namely, of
Eq. (3.1) if $\delta K$ satisfies
$$
\theta_i \delta K = \left( \delta L^i \right)_- K +
 \left( L^i \right)_- \delta K \eqno(3.4)
$$
One can show quite easily from the relations in Eqs. (2.13) and (3.2) that
$$
\delta\left( L^i \right)_- = \epsilon\left[ Q_- , L^i \right]_- \eqno(3.5)
$$
The symmetry condition, Eq. (3.4), can now be equivalently written as
$$
\theta_i Q_- = - \left[ L^i , Q_- \right]_-
+ \left[ \left( L^i \right)_-, Q_- \right] \eqno(3.6)
$$
This can be easily seen to take the simple form
$$
\theta_i Q_- = - \left[ \left( L^i \right)_+ , Q_- \right]_- \eqno(3.7)
$$
In obtaining this simple form, it was essential that the symmetry operator Q
is restricted to its
negative part in (3.2). It is often convenient to express the symmetry
generator as
$$
     Q=  K V K^{-1} \eqno(3.8)
$$
The symmetry condition, Eq. (3.7), can then be shown to be equivalent to
$$
\theta_i V = -\left[ \theta^i , V \right] \eqno(3.9)
$$
The construction of symmetries of the super-KP hierarchy is, then,
equivalent to determining $V$'s which satisfy Eq. (3.9).  As a familiar
example, let us note that
$$
V_n = \theta^{2n} \eqno(3.10)
$$
automatically satisfies the above relation. These are nothing other than the
familiar flows
generated by the bosonic conserved quantities of the theory.

However, we are interested in additional time dependent symmetries of the
theory.  Thus, we make an ansatz for V of the following form
$$
\eqalign{V= \alpha \xi &+ x (\beta \theta +
\gamma \xi \partial ) + \sum_{n \geq 1}
\big( \alpha_n^1 \theta + \alpha^2_n
\xi \partial \big) t_{2n} \partial^{n-1}\cr
\noalign{\vskip 4pt}%
&+ \sum_{n \geq 1} \big( \alpha^3_n + \alpha^4_n \xi \theta \big)
\tau_{2n-1} \partial^{n-1}\cr
\noalign{\vskip 4pt}%
&+ \sum_{n,k \geq 1} \big( \alpha^5_{nk} \theta +
\alpha^6_{nk} \xi \partial \big)
 \tau_{2n-1} \tau_{2k-1} \partial^{n+k-2}\cr}\eqno(3.11)
$$
where $\alpha, \beta, \gamma, \alpha^1_n, \dots \alpha^6_{nk}$ are constant
coefficients to be determined.  Requiring the symmetry condition to hold
for even flows, we obtain
$$
\eqalign{\alpha^1_n &= -\beta n\cr
\noalign{\vskip 4pt}%
\alpha^2_n &= - \gamma n\cr}\eqno(3.12)
$$
Similarly, requiring the symmetry condition to hold for the odd flows we
determine all but two constants.  Thus, a symmetry generator satisfying
Eq. (3.9) can be written as
$$
\eqalign{V= \alpha \xi &+ \beta x \overline \theta - \beta
\sum_{n \geq 1} n t_{2n} \partial^{n-1} \overline \theta +
 \sum_{n \geq 1} (-\alpha + (n-1) \beta) \tau_{2n-1}
\partial^{n-1}\cr
\noalign{\vskip 4pt}%
&- \beta \sum_{n \geq 1} (2n-1) \tau_{2n-1} \partial^{n-1}
 \xi \theta + \beta \sum_{n,k \geq 1} n \tau_{2n-1} \tau_{2k-1}
\partial^{n+k-2} \overline \theta \cr} \eqno(3.13)
$$
where $\alpha, \beta$ are the two arbitrary constants and
$$
\overline \theta = {\partial \over \partial \xi} -\xi \> {\partial \over
\partial x}\eqno(3.14)
$$
which satisfies
$$
\eqalign{[ \theta, \overline \theta ] &= 0\cr
\noalign{\vskip 4pt}%
[ \overline \theta, \overline \theta ] &= - 2 \partial \cr}\eqno(3.15)
$$
\bigskip
\noindent {\bf 4. Algebraic Structure of the Symmetry Transformations}
\bigskip
In this section, we would like to construct a canonical basis of the symmetry
operators which would lead to a natural (graded) Lie-algebraic structure.  From
experience, we realize that in the coordinate basis, it is most convenient to
isolate the symmetry operators into the form $x + \dots,
\> \xi+ \dots,\> \partial + \dots, \> {\partial \over \partial \xi} + \dots$.
 It
is trivially seen from the symmetry condition in Eq. (3.9) that the operator
$\partial = {\partial \over \partial x}$ is a symmetry operator.  Furthermore,
by choosing special values of the parameters $\alpha$ and $\beta$ in Eq. (3.13)
we obtain
$$
\eqalignno{\alpha = 1,\> \beta = 0 \> &: \qquad
V_1 = \xi - \sum_{n \geq 1} \tau_{2n-1} \partial^{n-1}\equiv T &(4.1)\cr
\alpha = 0,\> \beta = 1\> &: \qquad V_2 = X \overline \theta &(4.2)\cr}
$$
where
$$
\eqalign{ X = x - &\sum n t_{2n} \partial^{n-1} -
 {1 \over 2}\> \sum_{n \geq 1} \tau_{2n-1} \partial^{n-2} \big( (2n -1)
 \theta - \overline \theta \big)\cr
&+ \sum_{n,k \geq 1} n \tau_{2n-1} \tau_{2k-1} \partial^{n+k-2}\cr} \eqno(4.3)
$$
While T is already in the desired form, we need to reduce $V_2$ to the
canonical form.  To this end, we note that the commutator of two symmetry
operators is, itself, a symmetry operator.  Therefore, since
$$
\left[ \partial, V_2 \right] = \overline \theta \eqno(4.4)
$$
we conclude that $\overline \theta$ is a symmetry operator.  (In fact, it
generates space supersymmetry.)  Furthermore, from the structure of $V_2$,
which
is a symmetry operator, it is then obvious that $X$ must also be a symmetry
operator.  To find a symmetry operator with the coordinate form
 ${\partial \over \partial \xi} + \dots$, we note that
$$
   S\equiv  \overline \theta + T \partial = {\partial \over \partial \xi}\>
- \sum \tau_{2n-1} \partial^n \eqno(4.5)
$$
is also a symmetry operator.

{}From the structures of the symmetry operators $\partial,\> X,\>
S,\> T $
-- two of which are bosonic and the other two fermionic -- it is easy to
verify that
$$
[ \partial, X ] = 1 = [ S,T ]\eqno(4.6)
$$
with all other graded commutators vanishing.  One can, therefore, make the
following correspondence now, namely,
$$
\eqalign{X &\leftrightarrow z \qquad \partial \leftrightarrow {\partial
\over \partial z} \cr
\noalign{\vskip 4pt}%
T &\leftrightarrow \kappa \qquad S \leftrightarrow {\partial \over
\partial \kappa} \cr}\eqno(4.7)
$$
where $z$ is a bosonic variable while $\kappa$ is a Grassmann variable.  With
this identification, we can now show that the algebra of the symmetry operators
is nothing other than the super $W^+_\infty$ algebra.  We do this in the
next section.
\bigskip
\noindent {\bf 5. The Super $W_\infty $ Structure}
\bigskip
The super $W_\infty$ algebra was constructed in Ref. [13]. The bosonic
subalgebra is $W_\infty \oplus
W_{1+\infty}$ generated by $V^i_m$ and ${\tilde V}^i_m$, respectively, where
$i+2$ labels the
(quasi) conformal spin of the generator, and $-\infty\le m\le \infty$. For
$V^i_m,\
i=0,1,.. $ while for ${\tilde V}^i_m,\  i=-1,0,...$  The fermionic generators
are $G^{i\pm}_m$,
where the (quasi) conformal spin of the generator is now $i+{3\over 2},
\ i=0,1,...$ A realisation of
this algebra in terms of super differential operators was given in Ref. [14].
These operators are
as follows \footnote{$^\dagger$}{\tenfoot In Ref. [14], one  parameter
(denoted by $\lambda$ ) family
of super $W_\infty$ algebras are given, which correspond to different choices
of basis. Here we have
made the choice $\lambda=0$, which produces the super $W_\infty$ algebra
constructed explicitly in
Ref. [13]).}
$$
\eqalign{
  V^i_m &= \sum_{\ell=0}^{i+1}{1\over i+1}a^i_{m\ell}
\bigg[\ell+ (i+1-\ell)\kappa{\partial\over \partial \kappa} \bigg]
        z^{k-m}\bigg({\partial\over \partial z}\bigg)^\ell,  \cr
{\tilde V}^i_m &= \sum_{\ell=0}^{i+1}{1\over 2i+1}a^i_{m\ell}
\bigg[-\ell+(i+\ell+2)\kappa{\partial\over \partial\kappa}\bigg]
z^{k-m}\bigg({\partial\over \partial z}\bigg)^\ell,  \cr
G^{i\pm}_{m-{1\over 2}} &=\sum_{\ell=0}^i z^{\ell+1-m}\bigg({\partial\over
\partial z}\bigg)^\ell
\Bigg[{2\ell\over i+\ell+1}\ a^i_{m,\ell+1}\ \kappa{\partial\over \partial z}
\pm {i+\ell+1\over
2i+1}\ a^{i-1}_{m-1,\ell}\ {\partial\over \partial \kappa}\Bigg],   \cr}
\eqno(5.1)
$$
where
$$
    a^i_{m\ell} = \pmatrix{
i+1\cr \ell\cr}{(m-i-1)_{i+1-\ell}(-i-1)_{i+1-\ell}\over
(i+\ell+2)_{i+1-\ell}}, \eqno(5.2)
$$
where we have used the definition
$$
(a)_n\equiv {\Gamma(a+n)\over \Gamma(a)}=a(a+1)(a+2)\cdots (a+n-1), \quad
{\rm with}  \ \ (a)_0=1.
       \eqno(5.3)
$$

The reason for complicated choices for coefficients is to ensure that the
resulting generator has a
definite transformation property under the $SL(2,R)$ subalgebra of the
Virasoro algebra generated
by $V^0_m$. This $SL(2,R)$ covariance property allows one to assign
(quasi) conformal spin to each
genarator, thus facilitating the use of many conformal field theoretic
techniques in dealing with
this algebra.

It is clear now how to identify the super $W_{\infty}$ structure of the
symmetry transformations
of the previous section. Since any power of the basic symmetry operators
$\partial, X, S$ and $T$ is
also a symmetry operator, using the correspondence (4.7), we have the
following  symmetry
flows with manifest super $W_{\infty}$ structure:
$$
      \eqalign{
         {\partial\over \partial \varepsilon^i_m}K &= (KV^i_mK^{-1})_-K, \cr
{\partial\over \partial {\tilde \varepsilon}^i_m}K &=
(K{\tilde V}^i_mK^{-1})_-K, \cr
   {\partial\over \partial \varepsilon^{i\pm}_{m-{1\over 2}}}K &=
(KG^{i\pm}_{m-{1\over
2}}K^{-1})_-K,  \cr} \eqno(5.4)
$$
These are time dependent, non-isospectral flows which commute with the
isospectral super KP
flows (2.11). We emphasize that in (5.4), the replacements
$z\rightarrow X, \ {\partial\over \partial
z}\rightarrow \theta^2,\ \kappa\rightarrow T,\ {\partial\over \partial \kappa}
\rightarrow S $ are to
be made. For example, some of the low lying generators are given explicitly
as follows
$$
    \eqalign{
   V^0_m &=X^{1-m}\theta^2-{1\over 2}(m-1)X^{-m}TS, \cr
    V^1_m &=X^{2-m}\theta^4-{1\over2}(m-2)
(1+TS)X^{1-m}\theta^2 +{1\over 3}(m-1)(m-2)X^{-m}TS, \cr
 {\tilde V}^{-1}_m &=-X^{-m}TS, \cr
   {\tilde V}^0_m &=(-1+3TS)X^{1-m}\theta^2-(m-1)X^{-m}TS, \cr
  G^{0\pm}_{m-{1\over 2}} &= X^{1-m}(T\theta^2 \pm S), \cr
G^{1\pm}_{m-{1\over 2}} &= X^{2-m}\theta^2(T\theta^2 \pm S)
-{1\over 3}(m-2)X^{1-m}(2T\theta^2 \pm S). \cr}                \eqno(5.5)
$$
It is straightforward to show that $V^0_m$ obey the Virasoro algebra. Other
interesting subalgebras of the full super $W_\infty$ are discussed in
Ref. [13], and contraction down
to the classical version super $w_\infty$ which is equivalent to super
symplectic diffeomorphism of a
suitable supermanifold are discussed in Refs. [13] and [15]. In particular,
there exists the
subalgebra super $W_\infty^-$ where $m\le i+1$ in $V^i_m$, in which case no
negative powers of $z$
occur in (5.1).
\bigskip \noindent {\bf 6. Conclusions}
\bigskip
We have constructed flows (5.4) which commute with the Manin-Radul super KP
flows (2.11).  These, therefore, represent the symmetries of the Manin-Radul
super KP hierarchy.  We would like to emphasize here that the usual conserved
quantities associated with an integrable system also generate symmetries of the
system, but what we have obtained here are additional symmetries which are in
general
time dependent.

The action of the basic symmetry operators $\partial$, and $T, X, S $ defined
in (4.1), (4.3),
(4.5), respectively, on the vacuum solution (2.20)  can not all be
representated in terms of
differential operators involving the spectral parameters $\lambda$ and $\eta$
alone. In the case of
of the bosonic KP hierarchy this is possible. It may as well be possible, and
it would be useful, to
construct a similar representation of the symmetry operators in terms of the
spectral parameters in
the case of super KP hierarchy as well, by taking suitable combinations of
the symmetry operators
given in this paper.

We conclude by pointing out some further interesting problems which deserve
study. In Ref. [16] it
has been shown that the KP hierarchy has additional symmetries which obey a
Kac-Moody-Virasoro based
on a subalgebra of $Vir \oplus \hat s \ell (5,R)$.  It is not clear whether
this means that the
symmetry ring of the Manin-Radul system admits a Kac-Moody extended version
of $W^+_\infty$ which
contains the symmetry algebra of Ref. [16] as a subalgebra.  To understand
this, it will be useful
first to realize such an extended algebra in terms of differential operators
in a manner similar to
Eq. (5.1).

The KP hierarchy has other supersymmetric extensions as well [12]. It would be
interesting to explore the symmetry properties of these extensions in a
systematic manner and to compare with one another.  What would be very
interesting in this context is to find a theory of a relativistic membrane or
even higher extended objects whose equations of motion would belong to a KP
hierarchy, in which case the large symmetry of the hierarchy, such as those
found in this paper, could be utilized to generate infinitely many solutions,
and infinitely many relations between scattering matrix elements, thus
rendering
the theory, if not soluble, at least manageable.
\bigskip\bigskip
\centerline{\bf ACKNOWLEDMENTS}
\bigskip
 We would like to thank Professor Abdus Salam, the International Atomic
Energy Agency and UNESCO
for hospitality at the International Center for Theoretical Physics where
this work was done.
We would also like to thank M. G\"urses for bringing Ref. [16] into our
attention.
\vfill\eject

\centerline{\bf REFERENCES}
\bigskip

\item{1.}A.B. Zamolodchikov, Teo. Mat. Fiz. {\bf 65} (1985) 347;\ V.A. Fateev
and S. Lukyanov,  Int.
J. Mod.  Phys. {\bf A3} (1988) 507.
\item{2.} C.N. Pope, L.J. Romans and X. Shen, Phys. Lett. {\bf 236B} (1990)
173;\ Nucl. Phys. {\bf
B339} (1990) 191;\ Phys. Lett. {\bf 242B} (1990) 401.
\item{3.} M.A. Awada and S.J. Sin, {\it Twisted $W_\infty$ symmetry of the KP
hierarchy and the string equation of the d=1 matrix models}, preprint,
UFIFT-HEP-90-33.
\item{4.} M. Fukuma, H. Kawai and R. Nakayama, {\it Infinite dimensional
Grassmannian structure of
two dimensional gravity }, preprint, UT-572, KEK-TH-272 (1990);\ E. Witten,
{\it Ground ring of
two dimensional string theory }, preprint, UFTP-IASSNS-HEP-91/51;\
I. Klebanov and A. Polyakov,
{\it Interaction of discrete states in two dimensional string theory },
preprint, PUPT-181.
\item{5.} F. Magri, J. Math.Phys. {\bf 19} (1978) 1156;\ J.L. Gervais,
Phys. Lett. {\bf
160B} (1985) 277.
\item{6.} V. Drinfeld and V. Sokolov, J. Sov. Math. {\bf 30} (1985) 1975;\
A. Bilal and J.L. Gervais,
Phys. Lett. {\bf 206B} (1988) 412;\ I. Bakas, Nucl. Phys. {\bf B302} (1988)
189;\ A. Das and S. Roy,
Int. J. Modern Phys. {\bf 6} (1991) 1429.
\item{7.} A.M. Polyakov, Mod. Phys. Lett. {\bf A2} (1987) 893;\
V.Z. Knizhnik, A.M. Polyakov and A.B.
Zamolodhikov, Mod. Phys. Lett. {\bf A3} (1988) 819.
\item{8.} B.B. Kadomtsev and V.I. Petviashvili, Sov. Phys. Dokl. {\bf 15}
(1971) 539.
\item{9.} K. Yamagishi, Phys. Lett. {\bf 259B} (1991) 436;\ F. Yu and
Y.S. Wu, Phys. Lett. {\bf 263B}
(1991) 220;\ A. Das, W.J. Huang and S. Panda, {\it The Hamiltonian structures
of the KP
hierarchy }, University of Rochester preprint, UR-1219 (1991).
\item{10.} P.G. Grinevich and A.Yu. Orlov, {\it  Virasoro action on Riemann
surfaces,
Grassmannians, det ${\bar \partial}_J$ and Segal-Wilson $\tau$-function },
in  Problems in Modern Quantum Field Theory, Eds. A.A. Belavin, A.U. Klimyk
and A.B. Zamolodhikov
(Springer-Verag, 1989).
\item{11.} Yu. I. Manin and A.O. Radul, Commun. Math. Phys. {\bf 98} (1985) 65.
\item{12.} A. Yu. Orlov, {\it Vertex operator, ${\bar \partial}$--problem,
symmetries, variational
identities and Hamiltonian formalism for 2+1 integrable systems }, in  Plasma
Theory and
Non-linear and Turbulent  Processes in Physics, Eds. V.G. Bar'yakhtar,
V.M. Chernousenko, N.S.
Erokhin and V.E. Zakharov (World Scientific, 1988);\ V. Kac and
J. van de Leur, Ann. Inst. Fourier,
Grenoble {\bf 37} (1987) 99; A. LeClair, Nucl. Phys. {\bf B314} 425;\
M. Mulase, {\it A new KP
system and a characterization of the Jacobians of arbitrary algebraic super
curves }, ITD preprint,
88/90-9;\ J. Figueroa-O'Farrili and E. Ramos, preprint, KUL-TF-91/17.
\item{13.}E. Bergshoeff, C.N. Pope, L.J. Romans, E. Sezgin and X. Shen,
Phys. Lett. {\bf 245B}
(1990) 447.
\item{14.}E. Bergshoeff, B. de Wit and M. Vasiliev, Phys. Lett. {\bf 256B}
(1991) 199;\ {\it  The structure of the super $W_\infty(\lambda)$ algebra },
preprint, CERN
TH-6021-91. 
\item{15.} E. Sezgin, {\it  Aspects of $W_\infty$ symmetry }, in the
Proceedings of the 4th
Regional Conference on Mathematical Physics held in Tehran.
\item{16.} D. David, N. Kamran, D. Levi and P. Winternitz,
J. Math. Phys. {\bf 27} (1986) 1225.

\end